\documentclass[conference]{IEEEtran}

\usepackage[top=0.82in, bottom=1.03in, left=0.68in, right=0.68in]{geometry}
\usepackage[cmex10]{amsmath}
\DeclareMathOperator*{\argmin}{arg\,min} 

\usepackage{setspace}
\usepackage{graphicx}
\usepackage{subfigure}
\usepackage{tabularx}
\usepackage{arydshln}
\usepackage{umoline}
\usepackage{picinpar}
\usepackage{mathrsfs}
\usepackage{latexsym}
\usepackage{amssymb}
\usepackage{pifont}
\usepackage{amsfonts}
\usepackage{amsmath}
\usepackage{color}
\usepackage{fancyhdr}
\usepackage{float}
\usepackage{cases}
\usepackage{bm}
\usepackage{multirow}
\usepackage[english]{babel}
\usepackage{epsfig}
\usepackage{graphics}
\usepackage[Lenny]{fncychap}
\usepackage{rotfloat}
\usepackage{xcolor}
\usepackage{algorithm}
\usepackage{algpseudocode}

\usepackage{tabularx,booktabs}
\newcolumntype{C}{>{\centering\arraybackslash}X}

\ifCLASSINFOpdf
\else
\fi

\newcommand{\ignore}[1]{}

\newcommand{\bigzeroU}{{\lower2ex\hbox{\huge 0}}}
\newcommand{\bigzeroL}{{\lower-.1ex\hbox{\huge 0}}}

\def\Vec#1{\boldsymbol{#1}}
\newcommand\blfootnote[1]{%
  \begingroup
  \renewcommand\thefootnote{}\footnote{#1}%
  \addtocounter{footnote}{-1}%
  \endgroup
}

\begin{document}
\title{\vspace{-5mm}Power-Domain-Multiplexed Precoded Faster-Than-Nyquist Signaling for NOMA Downlink\vspace*{-3mm}} %
\author{
\IEEEauthorblockN{Prakash~Chaki}
\IEEEauthorblockA{Depart. of Info. \& Commun. Eng.\\
The University of Tokyo\\
E-mail: prakash@g.ecc.u-tokyo.ac.jp \vspace*{-7mm}}
\and
\IEEEauthorblockN{Shinya~Sugiura$^*$}
\IEEEauthorblockA{Institute of Industrial Science\\
The University of Tokyo\\
E-mail: sugiura@iis.u-tokyo.ac.jp \vspace*{-7mm}}
}
\markboth{}
{Shell \MakeLowercase{\textit{et al.}}: Bare Demo of IEEEtran.cls for Journals}
\maketitle
\begin{abstract}
In this paper, we propose a novel precoded faster-than-Nyquist (FTN) signaling scheme with power-domain-multiplexing in non-orthogonal multiple access (NOMA) downlink for tapping the joint benefits of high spectral efficiency and simultaneous multiuser connectivity. Non-orthogonality is introduced in both the symbol-interval (time) domain as well as in the multiple-access (power) domain to achieve a flexible resource allocation. Eigendecomposition of the FTN-specific intersymbol interference (ISI) matrix is used to achieve efficient cancellation of ISI, while successive interference cancellation is used to eliminate multiuser interference induced by NOMA. We derive the achievable analytical rate bound and demonstrate the numerical results of the bit error rate performance for the proposed scheme.
\end{abstract}
\begin{IEEEkeywords}
Eigenvalue decomposition, faster-than-Nyquist signaling, massive connectivity, non-orthogonal multiple access, power allocation, precoding, spectral efficiency, successive interference cancellation.
\end{IEEEkeywords}

\IEEEpeerreviewmaketitle

\section{Introduction}
\label{sec:intro}
%
\blfootnote{Preprint (accepted version) for publication in \textit{IEEE Global Communications Conference (IEEE GLOBECOM)}, 2022, DOI: 10.1109/GLOBECOM48099.2022.10001617. 
$\copyright$ 2022 IEEE. Personal use of this material is permitted. Permission from IEEE must be obtained for all other uses, in any current or future media, including reprinting/republishing this material for advertising or promotional purposes, creating new collective works, for resale or redistribution to servers or lists, or reuse of any copyrighted component of this work in other works.}
Orthogonal signaling transmission has been historically employed in communication systems owing to its tractability for demodulation.
Orthogonal signaling in the time domain, i.e., zero intersymbol interference (ISI), is guaranteed by the Nyquist criterion considered in pulse shaping.
Furthermore, in the multiple-access domain, orthogonal multiple access (OMA) schemes, such as frequency division multiple access, time division multiple access, code division multiple access, and orthogonal frequency division multiple access, ensured independent resource allocation to each user. 
With the growing demand for higher system capacity as well as massive connectivity in the next-generation communication standard, non-orthogonal signaling and multiple access techniques have gained increased research interest.
Faster-than-Nyquist (FTN) signaling~\cite{takumi-ftn-evolution} is a non-orthogonal modulation technique in the time domain, which employs a symbol interval shorter than that defined by Nyquist criterion at the expense of deliberately introduced ISI.
Also, non-orthogonal multiple access (NOMA)~\cite{saito_NOMAorig_2013} is a multiple access technique that is capable of flexibly assigning available bandwidth and time resources to users for accessing the wireless medium while deliberately introducing multiuser interference (MUI).

First studied in the late sixties and early seventies \cite{tufts1}--\nocite{tufts2}\cite{mazo}, FTN signaling is capable of transmitting $25\%$ faster than the rate of Nyquist signaling for sinc pulses without degrading error rate performance if the symbol interval is equal to $0.802$ times the Nyquist counterpart in the time domain. While Nyquist's zero-ISI criterion defines a symbol packing ratio $\tau=1$, FTN signaling typically uses $\tau<1$, which allows us to exploit the excess bandwidth unavailable by the Nyquist signaling with a realistic orthogonal shaping filter, such as root-raised cosine (RRC) filter.
Diverse detection schemes, including optimal maximum likelihood sequence estimation~\cite{liveris}, maximum a-posteriori detection~\cite{rusek}, frequency-domain equalization (FDE)~\cite{sugiura_fde}, and SVD-assisted precoding and decoding~\cite{takumi-svd-twc,takumi-ftn-evolution}, have been studied for eliminating ISI induced in FTN signaling. 

\begin{figure}[!t]
\centering
\includegraphics[scale=0.42]{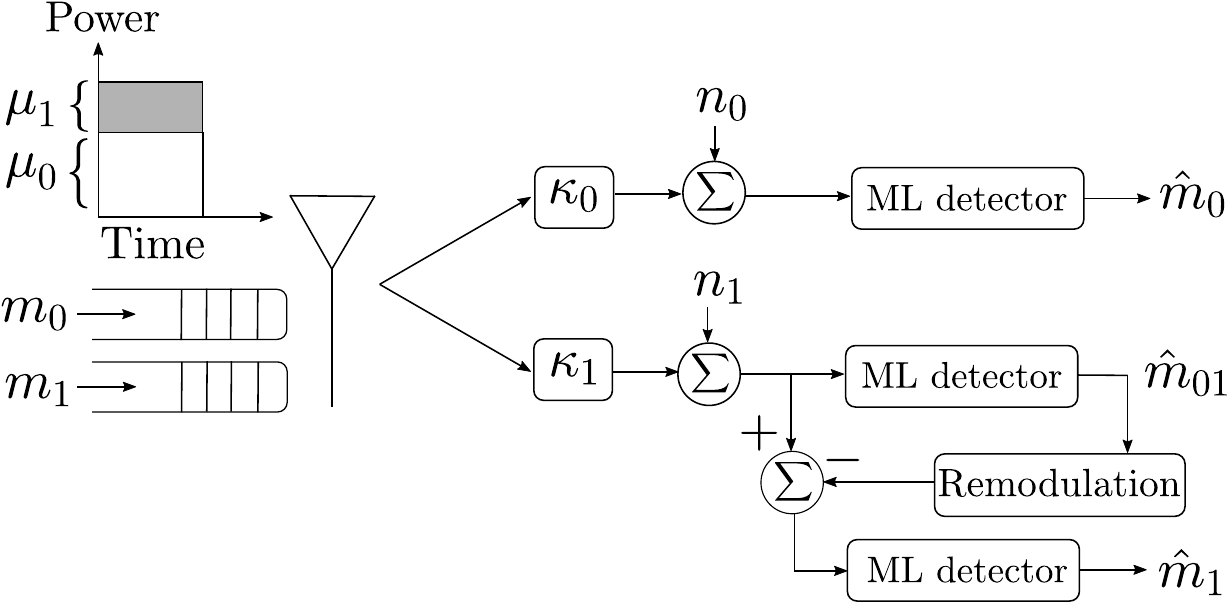}
\caption{Power-domain-multiplexed NOMA downlink for two users.}
\label{fig:illust}
\end{figure}

With the objective of massive connectivity, NOMA~\cite{ding_noma_appli_commag2017} supports multiple users in the overlapping time and frequency resources, i.e., in the power, the code, or the interleaver domains. 
For example, the power-domain NOMA downlink relies on superposition coding (SC) at the transmitter, where information blocks for multiple users are multiplexed in the power domain, as shown in Fig.~\ref{fig:illust}.
At the receiver, successive interference cancellation (SIC)\footnote{The combination of SC and SIC achieves the optimal performance in degraded broadcast channels \cite{tse_vishwanath_book}, \cite{cover}.} is used to recover information from the signals multiplexed in the power domain~\cite{saito_pimrc2013}.
Variants of NOMA have penetrated into the consideration for industry standards, e.g.,
LTE-A \cite{meredith2015study}, digital broadcasting~\cite{ldm_tb2016}, and massive machine type communication (mMTC) of 5G~\cite{ding_jsac2017}, \cite{dai_cst2018}.
Moreover, single-antenna single-cell NOMA~\cite{chen_tsp2017}, \cite{ding_spl2014}, multi-antenna single-cell NOMA~\cite{ding_twc2016}--\nocite{chen_tsp2017}\cite{zeng_jsac2017}, and multi-cell NOMA~\cite{you_twc2018}--\nocite{fu_twc2017}\cite{nguyen_jsac2017} have been investigated.

While the majority of FTN signaling was studied in the context of point-to-point OMA communication, FTN signaling in NOMA (FTN-NOMA) has the potential of providing higher spectral efficiency in the multiple-access scenario.
Existing FTN-NOMA studies were represented by \cite{ftn_noma_abebe_vtc2018}--\nocite{ftn_noma_yuan2020}\cite{ftn_noma_li2022}. More specifically, in \cite{ftn_noma_abebe_vtc2018}, the conventional FDE was introduced to FTN-NOMA, while in \cite{ftn_noma_yuan2020}, the iterative message-passing detector was developed for FTN-NOMA uplink with random access.
Furthermore, in \cite{ftn_noma_li2022}, the achievable rate of FTN-NOMA is investigated in the presence of random link delays of different users. To the best of our knowledge, the precoded FTN signaling in NOMA has not been investigated despite its potential.

Against the above background, the novel contribution of this paper is that we propose the power-domain-multiplexed precoded FTN signaling to enhance spectral efficiency in a NOMA downlink.
Specifically, the modulated symbols for each user are superposition encoded and then precoded with the aid of eigenvalue decomposition (EVD) prior to transmission at FTN rate. At the receiver of each user, the received signal is block-diagonalized to cancel out FTN-induced ISI, followed by MUI cancellation using SIC.
The rest of this paper is organized as follows. Section~\ref{proposed-system-model} presents the system model of our scheme, and the achievable ergodic rate is derived in Section~\ref{sec:ergodic_rate}. Furthermore, bit error ratio (BER) performance is provided in Section~\ref{sec:BER_two_user}. Finally, Section~\ref{sec:conclusion} concludes this paper.

\emph{Notations}: Boldface uppercase letters are used to denote matrices, boldface lowercase letters for vectors, and lowercase letters with a suffix to denote elements of a vector. $\mathbb{C}^{j\times k}$ and $\mathbb{R}^{j\times k}$ denote the complex and real fields of dimensions $j\times k$, respectively. {For complex values, $(.)^*$ denote conjugate operation.} For vectors or matrices, the transpose and Hermitian operations are denoted by $(.)^T$ and $(.)^H$, respectively. The expectation of a random variable is denoted by $\mathbb{E}[.]$.

\section{System Model of Proposed FTN-NOMA}
\label{proposed-system-model}
%
%
\begin{figure*}[!t]
\centering
\includegraphics[scale=0.38]{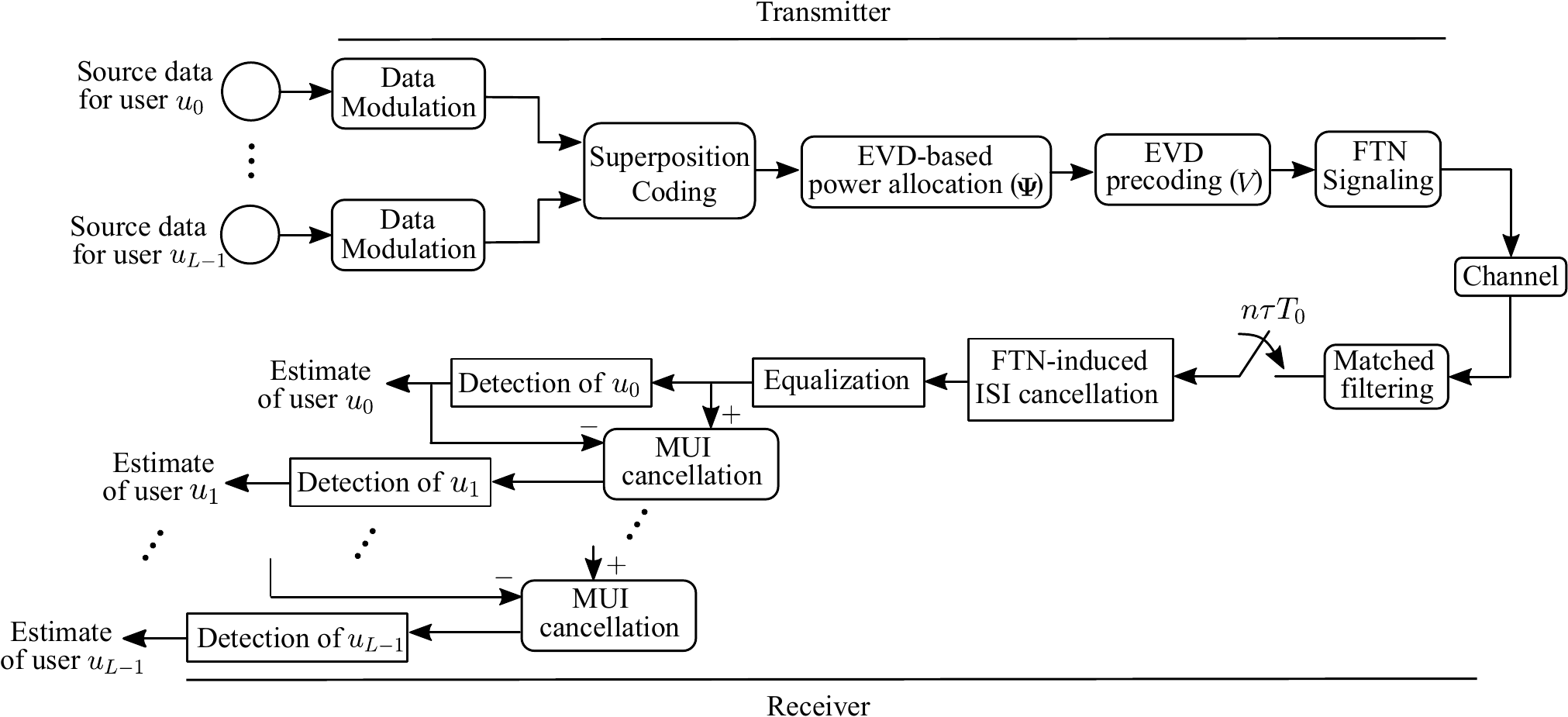}
\caption{Block diagram of proposed power-domain multiplexed precoded-FTN signaling scheme in $L$-user downlink.}
\label{fig:FTN_NOMA_block_dia}
\end{figure*}
Consider the power-domain-multiplexed precoded FTN-NOMA signaling system shown in Fig.~\ref{fig:FTN_NOMA_block_dia}, where a single BS and $L$ users $u_0,\cdots, u_{L-1}$ are assumed to be located in a single cell, each equipped with a single antenna.
Without loss of generality, we assume that the absolute values of the channel coefficients corresponding to $u_0,\cdots, u_{L-1}$ are in ascending order.
Information bits for the $l$th user are modulated onto $N$ complex-valued symbols, which are represented by $\Vec{s}_l=[s_{l,0}, s_{l,1}, \cdots, s_{l,N-1}]^T\in \mathbb{C}^N$. Here, the modulated symbols satisfies the power constraint of $\mathbb{E}[\Vec{s}_l^H\Vec{s}_l]=NP_s$, where $P_s$ is the average symbol power.
The modulated $L$ symbol vectors are then processed as follows. 
\subsection{Power-Domain Multiplexing}
\label{sec:system_model:superposition}
Based on the power-domain-multiplexing principle, the modulated symbols of the $l$th user are scaled by a power coefficient $\mu_l$, such that $\sum \mu_l=1$, which is referred to as power-domain-multiplexing-based power allocation (PDM-PA). 
The $\mu_l$ value is optimized in a dynamic manner based on knowledge of instantaneous channel state information (CSI)~\cite{yang_tcom2017}. For the sake of simplicity, we focus our attention on 
static power allocation, similar to~\cite{ding_tvt2016}. The PDM-PA symbols are multiplexed by superposition coding as follows:
\begin{align}
\Vec{c}&=[c_0, c_1, \cdots, c_{N-1}]^T\in\mathbb{C}^{N} \label{eq:superpos_code_3}\\
&=\sum_{l=0}^{L-1}\sqrt{\mu_lP_s}\Vec{s}_l. \label{eq:superpos_code_1}
\end{align}
\subsection{EVD-Based Precoding and Power Allocation}
\label{sec:system_model:precoding}
The superposition-coded symbol vector $\Vec{c}$ is multiplied by a real-valued diagonal matrix denoted by  $\Vec{\Psi}=\text{diag}(\sqrt{\psi_0},\sqrt{\psi_1},\cdots,\sqrt{\psi_{LM-1}})\in \mathbb{R}^{N\times N}$ as follows:
\begin{align}
\tilde{\Vec{c}}&=[\tilde{c}_{0}, \tilde{c}_{1}, \cdots, \tilde{c}_{N-1}]^T\in\mathbb{C}^{N}\\
&=\Vec{\Psi}\Vec{c}, \label{eq:EVD_PA}
\end{align}
which is referred to as EVD-based power allocation (EVD-PA).
The symbol vector $\tilde{\Vec{c}}$ is further multiplied by a complex-valued unitary matrix ${\bf V}\in\mathbb{C}^{N\times N}$ as follows:
\begin{align}
\Vec{x}&=[x_{0}, x_{1}, \cdots, x_{N-1}]^T\in\mathbb{C}^{N}\\
&={\bf V}\Vec{\tilde{c}}. \label{eq:EVD_precoding}
\end{align}
The matrices $\Vec{\Psi}$ and ${\bf V}$ are calculated by the EVD of the FTN-induced ISI matrix as explained in Sections~II-D and II-E, respectively.
Note that (\ref{eq:EVD_precoding}) corresponds to the EVD-based precoding block of Fig.~\ref{fig:FTN_NOMA_block_dia}.

\subsection{FTN Signaling}
\label{sec:system_model:signaling}
The precoded symbols of (\ref{eq:EVD_precoding}) are bandlimited by a pulse shaping filter with impulse response $h(t)$.
In this paper, we consider an RRC shaping filter with a roll-off factor $\beta$, which is $T_0$-orthogonal and satisfies the unit energy of $\int_{-\infty}^\infty\|h(t)\|^2dt=1$.
The symbol interval of our FTN-NOMA is $T=\tau T_0$, where the range of the packing ratio $\tau$ is defined by~\cite{ishihara2021techrxiv,sugiura2021twc}
\begin{align}
\frac{1}{1+\beta}\leq \tau \leq 1. \label{eq:tau_range}
\end{align}
The constraint (\ref{eq:tau_range}) allows us to avoid the scenario of a numerically ill-conditioned ISI matrix.

Then, the continuous-time transmit signal is generated as
\begin{equation}
x(t) = \sum_{n=0}^{N-1} x_n h(t-nT) \label{eq:transmit_sig}.
\end{equation}

\subsection{Matched Filtering and Sampling}
\label{sec:system_model:receiver:matched}
The signal received at the $l$th user is represented as follows
\begin{equation}
r_l(t) = \kappa_l \sum_{n=0}^{N-1} x_n h(t-nT) + \eta_l(t), \label{eq:received_sig}
\end{equation}
where $\eta_l(t)$ is a zero-mean complex-valued white Gaussian random process with a variance $N_0$.
Under the assumption of frequency-flat Rayleigh fading, {$\kappa_l=\sqrt{d_l^{-\zeta}}\Upsilon$ is the channel coefficient between the BS and $l$th user, where $\Upsilon$ is a complex normal random variable, $d_l$ is the distance between the BS and $u_l$, and $\zeta$ is the path loss exponent.}
As mentioned above, we assume the relationship of $|\kappa_0|<|\kappa_1|<\cdots<|\kappa_{L{-}1}|$. 

The received signal $r_l(t)$ is filtered by a matched filter $h^*(-t)$ as follows: 
\begin{equation}
\label{eq:match_fil_out}
r^{\text{MF}}_l(t) = \kappa_l\sum_{n=0}^{N-1} x_n g(t-nT) + \xi_l(t),
\end{equation}
where 
\begin{IEEEeqnarray}{rCL}
g(t)&=&\int{h(\varphi)h^*(t-\varphi)d\varphi}\\
\xi_l(t)&=&\int{\eta_l(\varphi)h^*(t-\varphi)d\varphi}.
\end{IEEEeqnarray}

By sampling the matched-filtered output with the FTN signaling interval of $\tau T_0$, we obtain the samples of
\begin{equation}
\label{eq:sampler_out}
r^{\text{MF}}_l(iT) = \kappa_l\sum_{n=0}^{N-1} x_n g(iT-nT) + \xi_l(iT).
\end{equation}
For ease of representation, (\ref{eq:sampler_out}) can be expressed in the vector form as
\begin{equation}
\label{recv_vec}
\Vec{r}^{\text{MF}}_l=\kappa_l{\bf G}\Vec{x}+\Vec{\xi}_l \in\mathbb{C}^{N}.
\end{equation}
Here, the noise components $\Vec{\xi}_l$ are correlated with the covariance matrix of
\begin{IEEEeqnarray}{rCL}
\mathbb{E}[\Vec{\xi}_{l}\Vec{\xi}_{l}^H]=N_0{\bf G}.
\end{IEEEeqnarray}
${\bf G}\in\mathbb{R}^{N\times N}$ is the ISI matrix resulting from non-orthogonality in symbol-interval dimension introduced by FTN signaling, whose $(u,v)$th element denoted by $g_{uv}$ captures the interference between the $u$th and $v$th symbols, $0\leq u,v<N$. 

To expound a little further, ${\bf G}$ is guaranteed to be full-rank 
\cite{kim_foldedspec_wcnc2016}.
Also, ${\bf G}$ is positive definite when (\ref{eq:tau_range}) is satisfied~\cite{takumi-ftn-evolution}.
Furthermore, $g_{uv}=g_{vu}$ for FTN signaling and $g_{uv}|_{u\neq v}=0$ for Nyquist signaling. The properties above ensure that ${\bf G}$ is a Toeplitz and Hermitian matrix. This enables us to perform its EVD-based factorization as follows:
\begin{equation}
{\bf G}={\bf V\Lambda V}^T,\label{eq:svd}
\end{equation}
where ${\bf \Lambda}=\text{diag}(\lambda_0, \lambda_1, \cdots, \lambda_{N-1})\in\mathbb{R}^{N\times N}$ contains the eigenvalues of ${\bf G}$ in its diagonal positions. The corresponding eigenvectors of ${\bf G}$ are captured in the orthonormal matrix ${\bf V}\in\mathbb{C}^{N\times N}$. The matrices ${\bf V}$ and ${\bf \Lambda}$ are used at the receiver to cancel ISI, as shown in Section~\ref{sec:system_model:ISI}.

\subsection{FTN-induced-ISI Cancellation}
\label{sec:system_model:ISI}
ISI is cancelled by diagonalization of the received samples $\Vec{r}^{\text{MF}}_l$ at the $l$th user. More specifically, by multiplying $\Vec{r}^{\text{MF}}_l$ with ${\bf V}^T$, we obtain the diagonalized received sample vector as follows:
\begin{align}
\Vec{r}^{\text{MF, diag}}_l&={[r^{\text{MF, diag}}_{l,0}, r^{\text{MF, diag}}_{l,1}, ...,r^{\text{MF, diag}}_{l,N-1}]^T\in\mathbb{C}^{N}}\\
&={\bf V}^T\Vec{r}^{\text{MF}}_l\in\mathbb{C}^{N}
\label{eq:diag1}\\
&=\kappa_l{\bf V}^T{\bf G}\Vec{x} + {\bf V}^T\Vec{\xi}_l\label{eq:diag2}\\
&=\kappa_l{\bf V}^T{\bf V\Lambda V}^T\Vec{x} + {\bf V}^T\Vec{\xi}_l\label{eq:diag3}\\
&=\kappa_l{\bf \Lambda V}^T{\bf V\Psi}\sum_{l=0}^{L-1}\sqrt{\mu_lP_s}\Vec{s}_l + {\bf V}^T\Vec{\xi}_l\label{eq:diag4}\\
&=\kappa_l{\bf \Lambda}\Vec{\Psi}\sum_{l=0}^{L-1}\sqrt{\mu_lP_s}\Vec{s}_l + {\bf V}^T\Vec{\xi}_l\label{eq:diag5}.
\end{align}
Here, (\ref{eq:diag3}) is obtained by substituting (\ref{eq:svd}) into (\ref{eq:diag2}),
and (\ref{eq:diag4}) is obtained by substituting (\ref{eq:EVD_precoding}) into (\ref{eq:diag3}).
Observe in (\ref{eq:diag5}) that the two terms on the r.h.s. are diagonal matrices, noting that 
${\bf V}^T\Vec{\xi}_l$ represents uncorrelated noise with the covariance matrix of $N_0\Vec{\Lambda}$.
Hence, we can carry out ISI-free symbol-by-symbol detection from (\ref{eq:diag5}).

The EVD-PA matrix $\Vec{\Psi}$ is optimized to maximize mutual information as follows~\cite{takumi-svd-twc}: 
\begin{equation}
\psi_n=\frac{1}{\lambda_n}, \hspace{2mm}0\leq n<N.\label{eq:optimal_EVD_PA}
\end{equation}
\subsection{MUI Cancellation and User-Wise Symbol Detection}
\label{sec:system_model:MUI}
The diagonalized vector $\Vec{r}^{\text{MF, diag}}_l$ is equalized to obtain $\Vec{r}^{\text{eq}}_l=[r^{\text{eq}}_{l,0}, r^{\text{eq}}_{l,1}, \cdots,r^{\text{eq}}_{l,N{-}1}]^T\in\mathbb{C}^{LM\times LM}$ as follows:
\begin{equation}
\label{eq:equalize}
r^{\text{eq}}_{l,n}=\frac{(\kappa_l\sqrt{\lambda_n})^*}{\lvert\kappa_l\sqrt{\lambda_n}\rvert^2+N_0\lambda_n} r^{\text{MF, diag}}_{l,n}.
\end{equation}
Subsequently, the signal of the user $u_0$ is demodulated from $\Vec{r}^{\text{eq}}_l$ in a symbol-by-symbol manner based on maximum likelihood (ML) detection as follows:
\begin{equation}
\hat{s}_n^{(0,l)}=\argmin_{s_{\theta}\in\Theta}\Big\lVert r^{\text{eq}}_{l,n}-s_{\theta}\Big\rVert^2, \label{eq:ML_user0}
\end{equation}
where $\hat{\Vec{s}}^{(0,l)}=[\hat{s}_0^{(0,l)}, \hat{s}_1^{(0,l)}, ..., \hat{s}_{N-1}^{(0,l)}]^T$ denotes the estimated symbol vector of $u_0$ detected at $u_l$, and $0\leq n<N$.
Also, $\Theta$ denotes legitimate symbol constellation space. 
At the user $u_l$, the signals of the users $u_1\cdots u_{l-1}$ are detected and removed in order. Then, the signal of the user $u_l$ given by  $\hat{\Vec{s}}^{(l,l)}=[\hat{s}_0^{(l,l)}, \hat{s}_1^{(l,l)}, \cdots, \hat{s}_{N-1}^{(l,l)}]^T$ is detected as follows 
\begin{equation}
\hat{s}_n^{(l,l)}=\argmin_{s_{\theta}\in\Theta}\bigg\lVert \big(r^{\text{eq}}_{l,n}-\sum_{z=0}^{l-1}\sqrt{\mu_zP_s}\hat{s}_n^{(z,l)} \big)-s_{\theta}\bigg\rVert^2, \label{eq:ML_userL}
\end{equation}
under the assumption $\mu_0>\mu_1>\cdots>\mu_{L{-}1}$.
%
%
\section{Achievable Ergodic Rate Analysis}
\label{sec:ergodic_rate}
%
%
\begin{figure}[!t]
\centering
\includegraphics[scale=0.55]{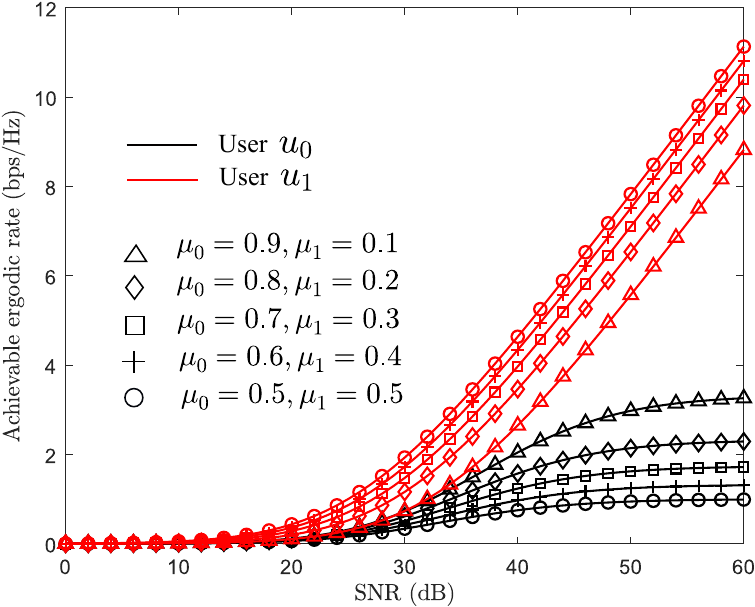}
\caption{Achievable ergodic rates in a two-user scenario of the proposed scheme under the condition of $\tau=1/(1+\beta)$.
}
\label{fig:achievable_rate}
\end{figure}
In this section, we demonstrate the achievable ergodic rate per user in the proposed power-domain-multiplexed precoded FTN signaling in NOMA downlink.

Consider the two-user setting introduced in Section~\ref{proposed-system-model}, where the transmit signal is denoted by (\ref{eq:superpos_code_1}).
The user $u_0$ has a lower channel gain of $\kappa_0$ and a PDM-PA coefficient $\mu_0$, while $u_1$ has a higher channel gain of $\kappa_1$ and a PDM-PA coefficient $\mu_1$.
Without loss of generality, we assume $\mu_0>\mu_1$.

The signal received at user $u_l, l\in\{0,1\}$ is expressed as
\begin{align}
\Vec{y}_l &= \kappa_l \Vec{c} + \Vec{\eta}_l, \label{eq:rec_sig_ul} \\
&= \kappa_l(\sqrt{\mu_0 P_s}\Vec{x}_0 + \sqrt{\mu_1 P_s}\Vec{x}_1) + \Vec{\eta}_l. \label{eq:rec_sig_ul_2}
\end{align}
The user $u_0$ demodulates $\Vec{x}_0$ with ML detection of $\Vec{y}_0$ since the term $\sqrt{\mu_0 P_s} \Vec{x}_0$ is dominant in $\Vec{y}_0$ by virtue of the relation $\mu_0>\mu_1$.
Here, the signal-to-interference-plus-noise ratio (SINR) associated with $u_0$ is characterized as
\begin{equation}
\rho_0 = \frac{|\kappa_0|^2\mu_0 P_s}{|\kappa_0|^2\mu_1 P_s+N_0}. \label{eq:sinr_u0}
\end{equation}
The user $u_1$ first estimates the signal $\Vec{x}_0$ by ML detection of $\Vec{y}_1$. Then, the term associated with $\Vec{x}_0$ is removed with the aid of SIC, and $\Vec{x}_1$ is demodulated by ML detection.
Assuming perfect SIC to eliminate MUI, the SINR of $u_1$ is expressed as
\begin{equation}
\rho_1 = \frac{|\kappa_1|^2\mu_1 P_s}{N_0}. \label{eq:sinr_u1}
\end{equation}
Furthermore, let us define the folded spectrum of the shaping pulse $h(t)$ as \cite{kim_foldedspec_wcnc2016}
\begin{equation}
H_{\text{FS}}(f) = \sum_{i=-\infty}^{\infty}\bigg\lvert H\bigg(f+\frac{i}{\tau T_0}\bigg)\bigg\rvert^2, \label{eq:folded_spec}
\end{equation}
where $H(f)$ is the frequency response of $h(t)$.
Note that $H_{\text{FS}}(f)$ has a non-zero value in the range of $-{1}/{\tau T_0} < f < {1}/{\tau T_0}$.
Assuming the transmission of Gaussian symbols, the maximum achievable ergodic rate 
of EVD-precoded FTN signaling with a pulse $h(t)$ in a point-to-point link is expressed by \cite{kim_foldedspec_wcnc2016} 
\begin{align}
R^{\text{EVD}} &= \int_{-1/(2\tau T_0)}^{1/(2\tau T_0)} \log_2\bigg(1+\frac{P_s}{N_0}H_{\text{FS}}(f)\bigg)
\end{align}

With (\ref{eq:EVD_PA}) and (\ref{eq:sinr_u0}), the maximum achievable ergodic rate at user $u_0$ is obtained as
\begin{align}
R_{u_0}^{\text{prop}} &= \lim_{N\rightarrow\infty}\frac{1}{N\tau T_0}\sum_{n=0}^{N-1}\log_2\bigg(1+\frac{|\kappa_0|^2\lambda_n\psi_nP_s\mu_0}{|\kappa_0|^2\lambda_n\psi_nP_s\mu_1 + N_0}\bigg). \label{eq:capacity_u0_eq1}
\end{align}
under the condition of (\ref{eq:tau_range}).
Note that (\ref{eq:capacity_u0_eq1}) is the sum of the capacities of $N$ independent substreams, which is enabled by the EVD-based diagonalization.

With (\ref{eq:optimal_EVD_PA}), (\ref{eq:capacity_u0_eq1}) is simplified in bits/sec to
\begin{align}
R_{u_0}^{\text{prop}} &= \frac{2W}{\tau}\log_2\bigg(1+\frac{|\kappa_0|^2 P_s\mu_0}{|\kappa_0|^2 P_s\mu_1 + N_0}\bigg). \label{eq:capacity_u0_eq2}
\end{align}
Similarly, using (\ref{eq:sinr_u1}), the achievable rate of the user $u_1$ in bits/sec is expressed as
\begin{align}
R_{u_1}^{\text{prop}} &= \frac{2W}{\tau}\log_2\bigg(1+\frac{|\kappa_1|^2 P_s\mu_1}{N_0}\bigg). \label{eq:capacity_u1}
\end{align}
For the packing ratio of $\tau=1/(1+\beta)$, the equality holds in (\ref{eq:tau_range}), and correspondingly we arrive at the condition of maximum degrees of freedom gain achievable under EVD-precoded FTN with EVD-PA. The corresponding maximum achievable rates for the users $u_0$ and $u_1$ in bits/sec are given by
\begin{IEEEeqnarray}{rCL}
R_{u_0}^{\text{prop}} &=& 2W(1+\beta)\log_2\bigg(1+\frac{|\kappa_0|^2 P_s\mu_0}{|\kappa_0|^2P_s\mu_1 + N_0}\bigg) \label{eq:capacity_u0_eq3} \\
R_{u_1}^{\text{prop}} &=& 2W(1+\beta)\log_2\bigg(1+\frac{|\kappa_1|^2 P_s\mu_1}{N_0}\bigg), \label{eq:capacity_u1_eq3}
\end{IEEEeqnarray}
respectively, where $W$ denotes the one-sided bandwidth of $h(t)$.
Observe that the achievable rates of (\ref{eq:capacity_u0_eq3}), (\ref{eq:capacity_u1_eq3}) 
match with those of Nyquist signaling bound with an ideal rectangular filter. We evaluate the maximum ergodic rates of the proposed scheme in a two-user scenario with frequency-flat Rayleigh fading in Fig.~\ref{fig:achievable_rate} for $\tau=1/(1+\beta)$, under different PDM-PA values. (\ref{eq:capacity_u0_eq3}) and (\ref{eq:capacity_u1_eq3}) are used after the normalization with the bandwidth to express the rate in units of bps/Hz. {Observe in Fig.~\ref{fig:achievable_rate} that the user $u_1$ has higher rates owing to the perfect removal of MUI. By allocating higher power to $u_0$, its rate improved at the cost of reduction in the rate of $u_1$.}
%
%
%
%
%
%
\section{BER Performance Results}
\label{sec:BER_two_user}
%
%
\begin{figure}[!t]
\centering
\includegraphics[scale=0.5]{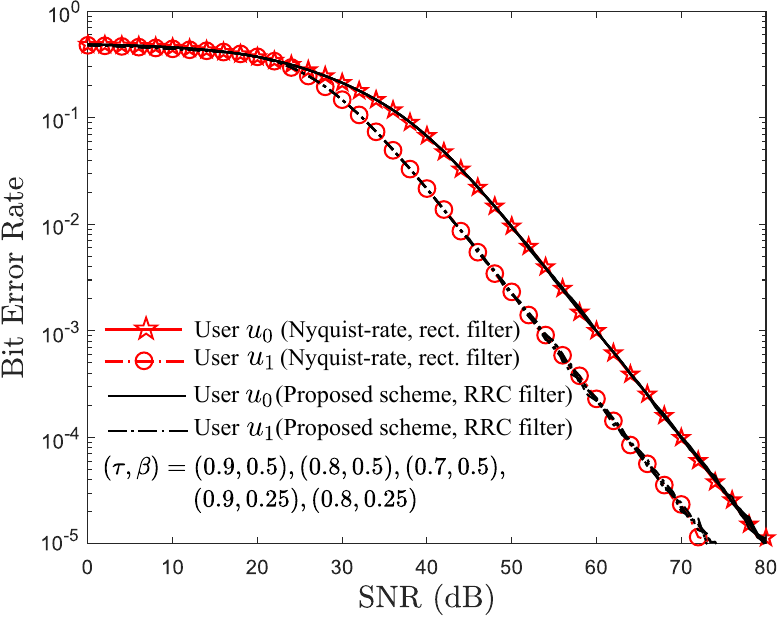}
\caption{BERs of the proposed scheme \textit{with} EVD-PA and the conventional Nyquist-signaling scheme. The ideal rectangular filter is employed for the Nyquist signaling scheme while the practical RRC filter is used in the proposed scheme. The parameters $(\tau,\beta)=(0.9,0.5), (0.8,0.5), (0.7,0.5), (0.9,0.25), (0.8,0.25)$ are employed for the proposed scheme.}
\label{fig:uncoded_AWGN_EVD_FTN_NOMA_a0p75_b0p25}
\vspace{-3mm}
\end{figure}
\begin{figure}[!t]
\centering
\includegraphics[scale=0.5]{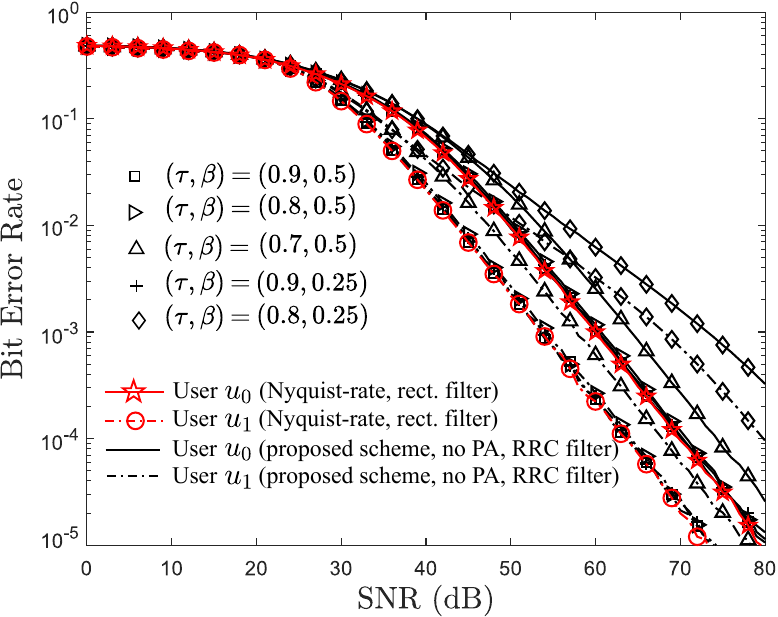}
\caption{BERs of the proposed scheme \textit{without} EVD-PA and the conventional Nyquist-rate-based scheme. The idea rectangular filter is employed for the Nyquist signaling scheme, while the practical RRC filter is used in the proposed scheme. The parameters $(\tau,\beta)=(0.9,0.5), (0.8,0.5), (0.7,0.5), (0.9,0.25), (0.8,0.25)$ are employed for the proposed scheme.}
\label{fig:uncoded_AWGN_EVD_FTN_NOMA_a0p75_b0p25_noPA}
\vspace{-3mm}
\end{figure}
\begin{figure}[!t]
\centering
\includegraphics[scale=0.5]{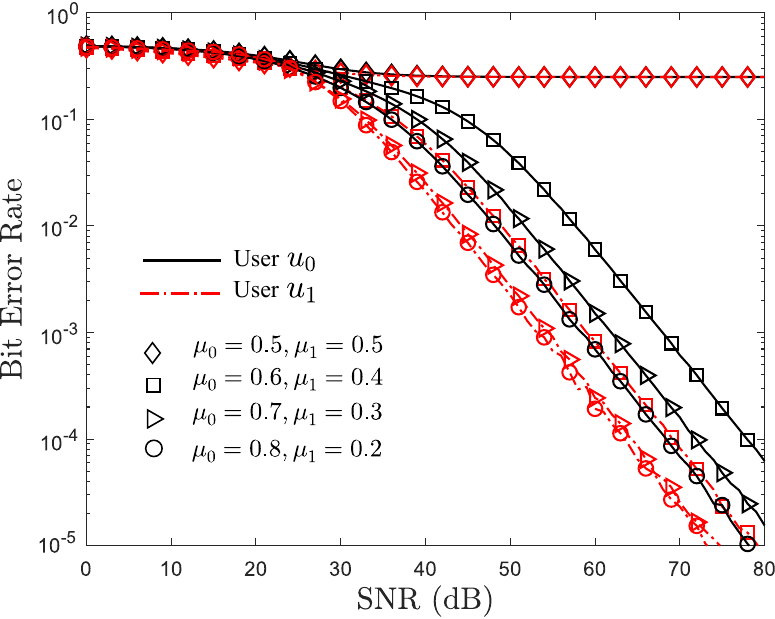}
\caption{BERs of the proposed precoded FTN-NOMA signaling with EVD-PA under different static power allocation ratio in PDM-PA between users. The parameters of $(\tau,\beta)=(0.8,0.25)$ are employed. User powers are given by $(\mu_1,\mu_2)=(0.5, 0.5), (0.6, 0.4), (0.8, 0.2), (0.9,0.1)$. The user powers are given by $(\mu_1, \mu_2)=(0.5, 0.5), (0.6, 0.4), (0.7, 0.3), (0.8, 0.2)$.}
\label{fig:uncoded_AWGN_EVD_FTN_NOMA_t0p8_b0p25_diff_NOMA_pow}
\vspace{-3mm}
\end{figure}
In this section, we present the numerical BER results in a channel-uncoded scenario. Fig.~\ref{fig:uncoded_AWGN_EVD_FTN_NOMA_a0p75_b0p25} shows the BER of the proposed precoded FTN-NOMA signaling in a two-user downlink, {where the distance of each user from the BS is given by $d_0=10$ m and $d_1=5$ m, respectively. Path loss exponent is set to $\zeta=3$.} 
The proposed scheme employed a practical RRC filter with a roll-off factor $\beta$. The parameters of $(\tau,\beta)=(0.9,0.5), (0.8,0.5), (0.7,0.5), (0.9,0.25), (0.8,0.25)$ are used. Furthermore, the BER of the conventional NOMA with Nyquist signaling and the ideal rectangular filter is also shown for comparison.
PDM-PA parameters of both the schemes are given by $(\mu_1, \mu_2)=(0.75, 0.25)$ and the transmission rate is $1$ bpcu.
Observe in Fig.~\ref{fig:uncoded_AWGN_EVD_FTN_NOMA_a0p75_b0p25} that regardless of the $\tau$ and $\beta$ values, the BER performance of the proposed scheme matches that of the Nyquist-signaling bound with the ideal rectangular filter.
Thus, the proposed scheme achieves the $1/\tau$ times higher bandwidth efficiency than the Nyquist counterpart with the same practical RRC shaping filter.
This is attained as a benefit of our efficient ISI cancellation procedure. 
{The user $u_1$ has better BER performance than the user $u_0$ owing to perfect SIC.}

Next, Fig.~\ref{fig:uncoded_AWGN_EVD_FTN_NOMA_a0p75_b0p25_noPA} shows the BER of the proposed scheme without EVD-PA, which corresponds to $\Vec{\Psi} = \mathbf{I}$, instead of employing (\ref{eq:optimal_EVD_PA}). Other parameters are the same as those employed in Fig.~\ref{fig:uncoded_AWGN_EVD_FTN_NOMA_a0p75_b0p25}.
As evident from Fig.~\ref{fig:uncoded_AWGN_EVD_FTN_NOMA_a0p75_b0p25_noPA}, ISI may not be successfully cancelled without EVD-PA, while it was with EVD-PA in Fig.~\ref{fig:uncoded_AWGN_EVD_FTN_NOMA_a0p75_b0p25}.

Furthermore, in Fig.~\ref{fig:uncoded_AWGN_EVD_FTN_NOMA_t0p8_b0p25_diff_NOMA_pow}, we show the impact of the PDM-PA coefficients.
The symbol packing ratio $\tau$ is set to $0.8$, and an RRC filter with roll-off $\beta=0.25$ is employed.
Also, the static PDM-PA parameters $(\mu_1,\mu_2)=(0.5, 0.5), (0.6, 0.4), (0.7, 0.3)$ and $(0.8, 0.2)$ are employed.
Typically, SIC works when the two users have a high power difference, which is also observed in Fig.~\ref{fig:uncoded_AWGN_EVD_FTN_NOMA_t0p8_b0p25_diff_NOMA_pow}.
For $(\mu_1, \mu_2)=(0.5, 0.5)$, there is no power difference between the two users, and hence SIC did not successfully eliminate the effects of MUI, resulting in the presence of an error floor.
Moreover, upon increasing
the power difference between the two users as $(\mu_1, \mu_2)=(0.6, 0.4)$, $(\mu_1, \mu_2)=(0.7, 0.3)$, and $(\mu_1, \mu_2)=(0.8, 0.2)$, the BERs of the two users are found to improve. 
%
%
\section{Conclusions}
\label{sec:conclusion}
%
In this paper, we proposed the precoded-FTN signaling scheme for NOMA downlink.
The proposed scheme aims at serving the dual purposes of achieving high spectral efficiency and simultaneous multiuser connectivity.
We provided the analysis of the per-user ergodic rate achievable in the proposed scheme. We demonstrated that in the BER performance for a two-user scenario, the achievable Nyquist-signaling bound with the ideal rectangular filter is achieved in the proposed scheme with a practical RRC shaping filter. %
\section*{Acknowledgement}
This study was supported in part by JSPS KAKENHI (Grant 22H01481), in part by JST PRESTO (Grant JPMJPR1933), and in part by JST FOREST (Grant JPMJFR2127).

\bibliographystyle{IEEEtran}
\bibliography{mybibfile}

\end{document}